\documentclass[]{spie}

\usepackage{amsmath,amsfonts,amssymb}
\usepackage{graphicx}
\usepackage[colorlinks=true, allcolors=blue]{hyperref}
\usepackage{authblk}

\title{Assembly development for the Simons Observatory focal plane readout module}

\author[a]{Erin Healy}
\author[b]{Aamir M. Ali}
\author[c]{Kam Arnold}
\author[d]{Jason E. Austermann}
\author[d]{James A. Beall}
\author[a]{Sarah Marie Bruno}
\author[e,f]{Steve K. Choi}
\author[d,g]{Jake Connors}
\author[h]{Nicholas F. Cothard}
\author[d,g]{Bradley Dober}
\author[d]{Shannon M. Duff}
\author[c]{Nicholas Galitzki}
\author[d]{Gene Hilton}
\author[i]{Shuay-Pwu Patty Ho}
\author[d]{Johannes Hubmayr}
\author[j]{Bradley R. Johnson}
\author[e]{Yaqiong Li}
\author[d]{Michael J. Link}
\author[d]{Tammy J. Lucas}
\author[a]{Heather McCarrick}
\author[e,f]{Michael D. Niemack}
\author[c]{Maximiliano Silva-Feaver}
\author[a]{Rita F. Sonka}
\author[a]{Suzanne Staggs}
\author[e]{Eve M. Vavagiakis}
\author[d]{Michael R. Vissers}
\author[a]{Yuhan Wang}
\author[b]{Benjamin Westbrook}
\author[h]{Edward J. Wollack}
\author[l,m]{Zhilei Xu}
\author[a]{Kaiwen Zheng}

\affil[a]{Department of Physics, Princeton University, Princeton, NJ, USA}
\affil[b]{Department of Physics, University of California, Berkeley, CA, USA}
\affil[c]{Department of Physics, University of California San Diego, La Jolla, CA, USA}
\affil[d]{National Institute of Standards and Technology, Boulder, CO, USA}
\affil[e]{Department of Physics, Cornell University, Ithaca, NY, USA}
\affil[f]{Department of Astronomy, Cornell University, Ithaca, NY, USA}
\affil[g]{Department of Physics, University of Colorado Boulder, Boulder, CO, USA}
\affil[h]{Department of Applied and Engineering Physics, Cornell University, Ithaca, NY, USA}
\affil[i]{Department of Physics, Stanford University, CA, USA}
\affil[j]{University of Virginia, Department of Astronomy, Charlottesville, VA, USA}
\affil[k]{NASA/Goddard Space Flight Center, Greenbelt, MD, USA}
\affil[l]{Department of Physics and Astronomy, University of Pennsylvania, Philadelphia, PA, USA}
\affil[m]{MIT Kavli Institute, Massachusetts Institute of Technology, Cambridge, MA, USA}

\authorinfo{E. Healy: E-mail: ehealy@princeton.edu}
\begin{document}

\maketitle
\begin{abstract}
The Simons Observatory (SO) is a suite of instruments sensitive to temperature and polarization of the cosmic microwave background (CMB) to be located at Cerro Toco in the Atacama Desert in Chile. Five telescopes, one large aperture telescope and four small aperture telescopes, will host roughly 70,000 highly multiplexed transition edge sensor (TES) detectors operated at 100\,mK. Each SO focal plane module (UFM) couples 1,764 TESes to microwave resonators in a microwave multiplexing ($\mu$Mux) readout circuit. Before detector integration, the 100\,mK $\mu$Mux components are packaged into multiplexing modules (UMMs), which are independently validated to ensure they meet SO performance specifications. Here we present the assembly developments of these UMM readout packages for mid frequency (90/150\,GHz) and ultra high frequency (220/280\,GHz) UFMs.

\end{abstract}


\section{Introduction} \label{intro}
\subsection{Simons Observatory}

In the last three decades since the absolute temperature\cite{cobe} and anisotropies of the CMB were first mapped\cite{Smoot}, increasingly sophisticated ground-based instruments, such as BICEP/Keck, SPT, ACT, and POLARBEAR \cite{BICEP, SPT, ACT, POLARBEAR}, have contributed greatly to our understanding of the universe and its evolution. The Simons Observatory's high precision CMB imagers, which will be located in the Atacama Desert in Chile, will significantly advance the next generation of cosmological discovery.

To achieve its science goals, SO will deploy a large aperture telescope (LAT) with a 6\,m primary mirror and four 42\,cm refracting small aperture telescopes (SATs) \cite{Galitzki}. This combination of angular scales will enable SO to measure the CMB anisotropies from a few degrees to acrminute scale. The LAT will observe approximately 40\% of the sky and the SATs approximately 10\%. The SATs are primarily designed to constrain the tensor-to-scalar ratio \emph{r} by measuring large-scale B-mode (divergence-free) polarization. The LAT will measure the temperature and polarization power spectra of the CMB, which will contribute to many science targets. These targets include the sum of neutrino masses, effective number of relativistic species, and other deviations from $\Lambda$CDM. With a large sky coverage, data from the LAT will be used in combination with other surveys, including DESI and LSST \cite{DESI, LSST}, for further cosmological and astrophysical analysis\cite{SO}.

The LAT will have 13 optics tubes, each of which can contain three detector arrays, with a total field of view of $\sim$7.8$^{\circ}$ in diameter. SO plans to populate seven of the thirteen optics tubes. The SATs will each have a single optics tube that can house up to seven arrays with a rotating cryogenic half-wave plate \cite{HWP} to modulate the low-frequency atmosphere signal. In total, SO plans to deploy $\sim$70,000 superconducting transition edge sensor (TES) detectors \cite{Lee, irwin05} across the five instruments. Each pixel, which consists of four TESes, will be sensitive to two orthogonal polarizations and two frequencies. SO will observe in six frequency bands. The low frequency (LF) detectors will measure in two bands centered at 30 and 40\,GHz. The mid frequency (MF) bands will be centered at 90 and 150\,GHz and the ultra-high frequency (UHF) detectors at 220 and 280\,GHz.

\subsection{Microwave multiplexing} \label{umux}

\begin{figure} [t]
   \begin{center}
   \begin{tabular}{c} 
   \includegraphics[height=7.8cm]{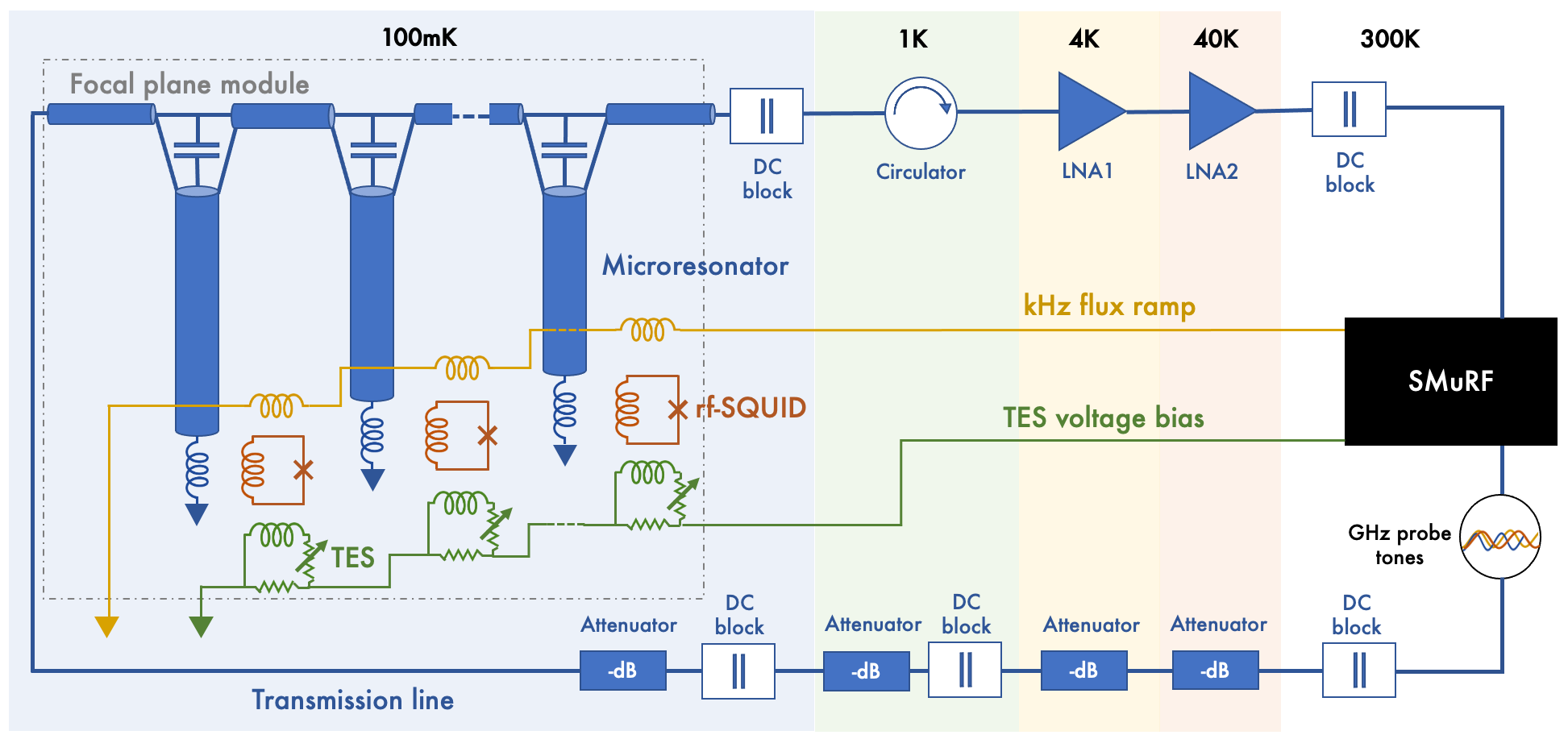}
   \end{tabular}
   \end{center}
   \caption[example] 
   { \label{fig:umux_ufm_smurf} 
Schematic of the $\mu$Mux readout circuit for the Simons Observatory. Resonators with unique frequencies between 4 and 6\,GHz are capacitively coupled to a common transmission line. Each resonator is inductively coupled to an rf-SQUID, which is inductively coupled to both a common flux ramp modulation line and a TES detector channel. The TESes are biased in series. The dotted gray box encloses all components housed in the SO focal plane modules. SLAC microresonator radio frequency (SMuRF) electronics generate the microwave probe tones, as well as the TES bias and flux ramp. When a photon signal transduces a shift in the resonance of a channel, the RF signal is amplified at two cryogenic stages (4\,K and 40\,K) and demodulated with SMuRF.\cite{Henderson18,Rao} }
   \end{figure}

A key path to improving sensitivity for CMB instruments is increasing the multiplexing factor, which enables a larger number of detectors to be read out on a smaller number of wires. Multiplexed readout reduces wiring costs and complexity, and it is especially important for low-temperature detectors, where the savings in thermal loading on the focal plane can be significant. There are several proven multiplexing architectures that amplify the detected photon signal using superconducting quantum interference devices (SQUIDs)\cite{ahmed, Henderson16, bender, suzuki}. In recent years, microwave SQUID multiplexing\cite{irwin04, Mates08} ($\mu$Mux) has emerged as a promising candidate for unlocking highly densely packed focal planes in CMB instruments\cite{dober}. Microwave multiplexing involves a large number of superconducting microwave resonators, each of which encodes a TES's response at a unique frequency, fabricated on a single silicon chip. These spectrally non-overlapping signals share a common transmission line, which is cryogenically amplified and carried to room temperature electronics with coaxial cables. The high multiplexing factor targeted by SO is $\mathcal{O}$(1000), larger than any other CMB instrument to date. Details of the SO multiplexing circuit can be seen in Figure \ref{fig:umux_ufm_smurf}.

\subsection{The universal focal plane modules}

The SO detector modules, the universal focal plane modules (UFMs), are hexagonal packages that house the 100\,mK $\mu$Mux components and the detector array. The package is designed for dense tiling of the detectors. The readout components are stacked behind the detectors, occupying no focal plane area. There are three varieties of UFMs, corresponding to the three frequency bands (LF, MF, UHF). In all three varieties, AlMn TES bolometers\cite{Deiker} with T$_{c}\sim$160\,mK are fabricated on 150\,mm silicon wafers. Though the details of the packaging differ between the three UFM varieties, they share the same basic design: a vertical stack of detector and $\mu$Mux readout components that are electrically connected via wire bonds in a package that can interface to both types of optics tubes (LAT and SAT). In total, SO plans to deploy 49 UFMs. These proceedings focus on the assembly of the UFM readout components of the MF and UHF arrays, which are identical. 



\section{Universal multiplexer module design} \label{design}

The 100\,mK readout components have their own packaging, called the universal multiplexing module (UMM), which is mated to the detector array in a UFM, as described in Section \ref{UFM_integration}. The UMM design detailed here was developed to improve the RF environment for the microwave resonators as compared to the prototype described in Ref.~\citenum{Li}. In this version, the multiplexer (mux) chips, on which the resonators and SQUIDs are fabricated, are sandwiched between two grounded copper plates. The plates provide a nearby RF ground plane to which the mux chips can be wire bonded. This packaging also creates a box around rows of mux chips, designed to reduce loss due to box mode coupling to the package, a standard loss mechanism for microwave resonators\cite{Sage}.

\subsection{Silicon components} \label{Si-parts}
At the core of the $\mu$Mux architecture are the multiplexer chips, fabricated at NIST. Each 4$\times$20\,mm$^2$ chip houses 64 readout circuits that can couple to TESes\cite{dober2020microwave}. The readout circuit for each channel, as described in Section \ref{umux}, consists of a microwave resonator inductively coupled to an rf-SQUID, which is inductively coupled to the flux ramp as well as the TES channel. The 64 channels are capacitively coupled to a coplanar waveguide (CPW) transmission line. The UMM is electrically split into two halves, called ``north" and ``south", with two symmetric CPW circuits that share a common RF ground. Each north/south half of the UMM consists of 14 mux chips in series for a total of 28 mux chips per UMM.

The routing wafer, fabricated on a 150\,mm silicon wafer at NIST, contains the TES bias circuity (including shunt resistors), the CPW transmission lines, and flux ramp lines. The twelve bias lines branch into individual channel circuits, routing from perimeter bond pads (for connection to TES detector wafer) to corresponding resonator channels on the mux chips. All electrical lines on the routing wafer are niobium, and they are connected to the detector wafer and mux chips with 25\,$\mu$m diameter aluminum wire bonds. The routing wafer also has windows for the mux chips and PCBs. On the east side of the wafer is a cutout window for the microwave launch (RF) PCB; on the west side is a cutout for the DC PCB, from which the bias lines and flux ramps are routed.

   \begin{figure} [t]
   \begin{center}
   \begin{tabular}{c} 
   \includegraphics[height=6cm]{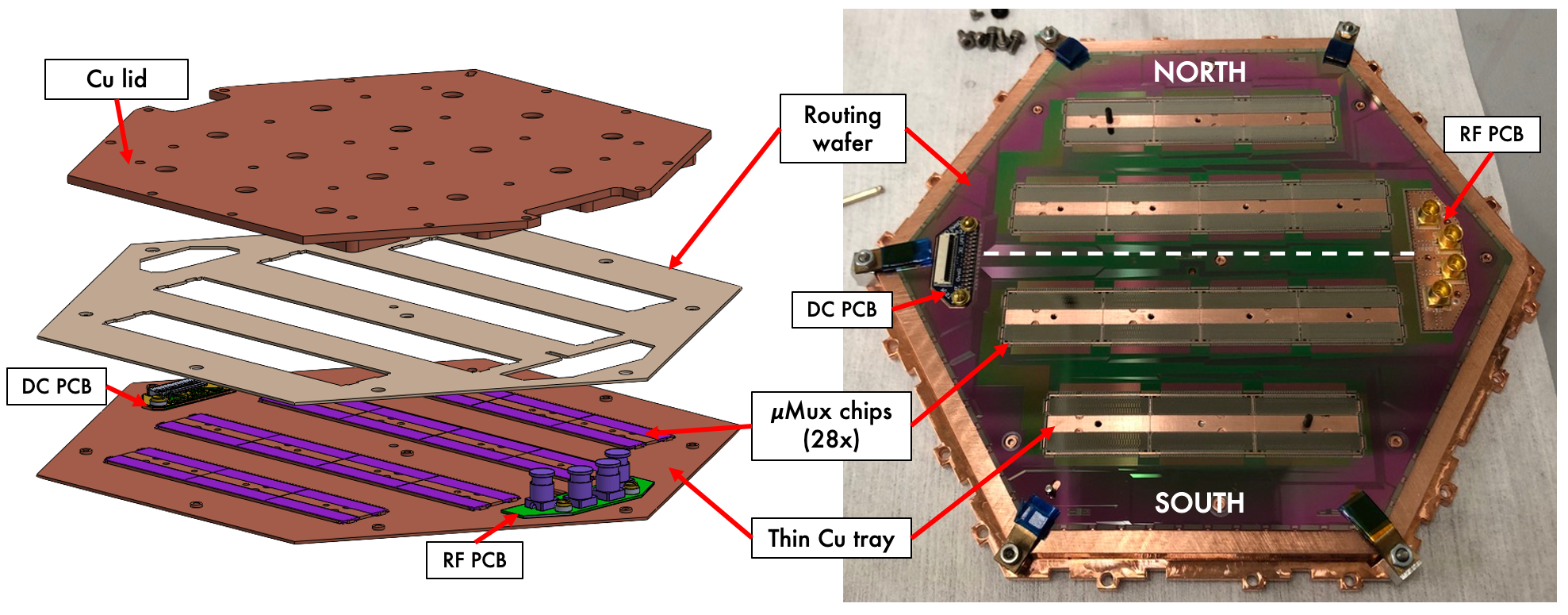}
   \end{tabular}
   \end{center}
   \caption[example] 
   { \label{fig:UMM_open} 
   \textbf{Left}: Exploded rendering of the UMM assembly. \textbf{Right}: Photo of an assembled UMM without the lid. Twenty-eight multiplexer chips are assembled in series on a thin copper tray, totaling 1,800 channels. Electrical lines fabricated on the routing wafer connect the RF and flux ramp signal to the mux chips via aluminum wire bonds. There are two multiplexed circuits on each UMM, labeled north and south. The PCBs interface to optics tube readout wiring.
}
   \end{figure}

\subsection{Other components}
The mechanical package consists of a thin (500\,$\mu$m) copper tray and a mating lid. All silicon parts and PCBs are mounted on the copper tray. The tray thickness is constrained by the UFM coupling. Aluminum wire bonds electrically connect the routing wafer to the detector wafer, so the vertical distance between these two wafers must not exceed 2\,mm to ensure strong wire bonds.

The tray has extruded features to which the mux chips and the routing wafer are grounded with wire bonds. The mux chips and routing wafer sit on the thin tray such that all bonding surfaces are coplanar. This improves conditions for bonding as well as minimizes the wire bond length, which is critical for reducing impedance mismatches along the transmission line. Due to the difference in thermal contractions between silicon and copper, the extruded features of the copper tray must be machined with tight (25---50\,$\mu$m) tolerances. The copper lid bolts directly to the thin tray along mating surfaces between mux chip rows. Spring-loaded contact pins glued into the lid provide clamping force on the routing wafer. There are two PCBs that feed the electrical lines to the optics tube, one for the RF signal and the other for the DC (TES bias) and low-frequency signal (flux ramp). A photo of an assembled UMM without the lid is in Figure \ref{fig:UMM_open}.

\section{UMM assembly}

The assembly procedure has been developed to ensure minimal degradation of the multiplexer performance due to assembly. This will enable the UMMs to the meet requisite performance specifications: a noise level of 65\,$\frac{pA}{\sqrt{Hz}}$ and multiplexing factor of 882 (not all resonators are coupled to detectors). It is critical that the assembly procedures are reliable and repeatable, as SO plans to field 49 detector modules. The devices must also be mechanically robust to handling and electrically functional (no electrical shorts or opens). The major assembly steps include: screening of components, aligning and clamping the routing wafer to the thin tray, gluing of multiplexer chips to the thin tray,  wire bonding, electrical validation, and final packaging.

\subsection{Component screening} \label{intake}
All components are visually inspected for defects and checked for mechanical fit before continuing with assembly. Silicon components undergo additional electrical screening before they are accepted. The mux chips are individually electrically validated with a custom probe card at room temperature that checks for continuity and shorts to ground on the flux ramp and CPW lines. The resonator performance must also be proven before proceeding with assembly. As mux chip performance is largely uniform across a fabrication wafer, one sample mux chip per wafer is packaged for cryogenic testing. The testing checks that the mux chips meet the design specifications for resonator parameters, including the internal quality factor (Q$_{i}$) and yield.

\begin{figure} [t]
   \begin{center}
   \begin{tabular}{c} 
   \includegraphics[height=7cm]{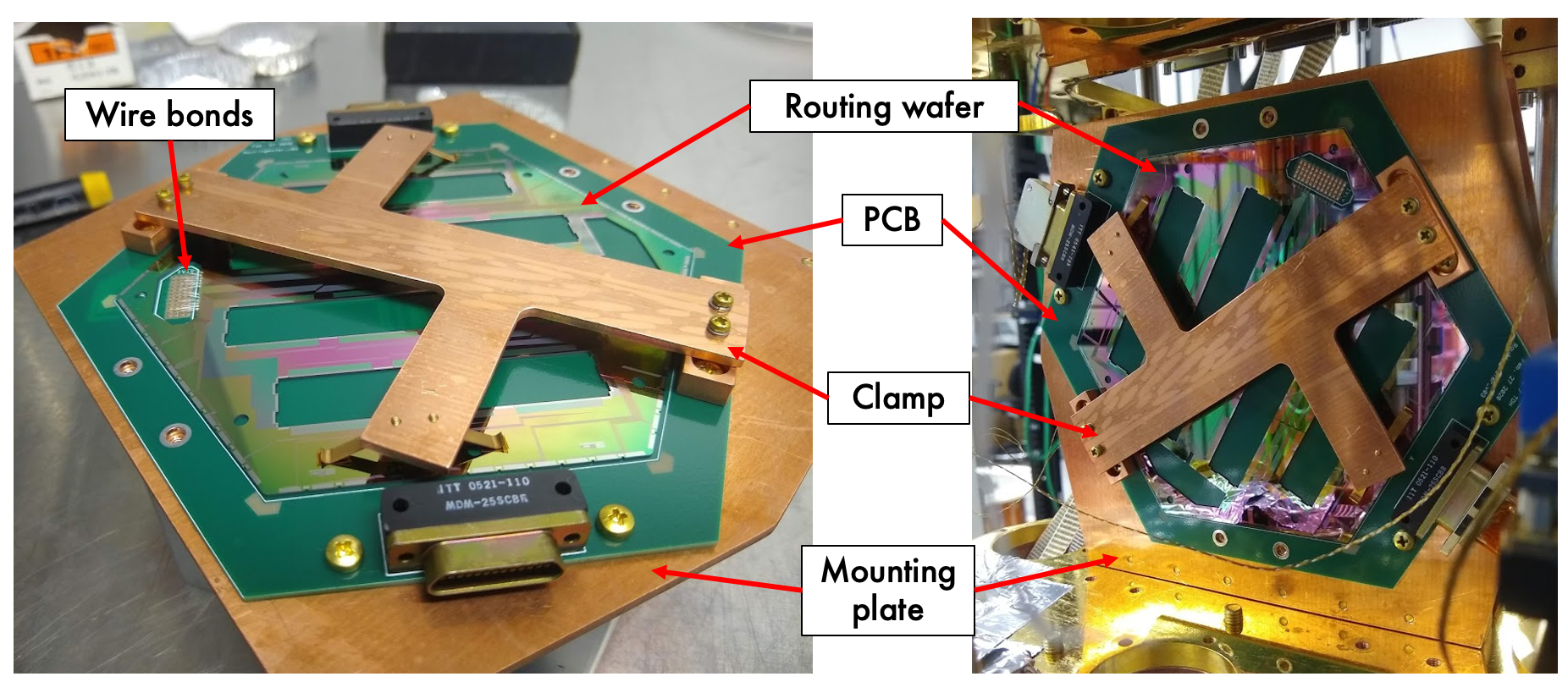}
   \end{tabular}
   \end{center}
   \caption[example] 
   { \label{fig:wafer_screener} 
   Photos from a routing wafer screening. \textbf{Left}: The wafer is placed on a PCB, and the bias lines on the wafer are wire bonded to the PCB for four-lead measurements. \textbf{Right}: A clamp holds the wafer in place against the PCB and mounting plate during cryocycle to 100\,mK. When cold, the bias line resistance and critical current of each bias line is measured, as well as resistance between bias lines.
}
   \end{figure}

Similarly, the routing wafer undergoes warm and cold screening. Following a visual inspection for defects, all critical electrical lines (bias lines, CPW lines, flux ramps) on the wafer are probed for continuity and shorts at room temperature using a current-controlled probing box. The wafer is then assembled for cryogenic screening at 100\,mK. This involves mounting the wafer to a PCB and wire bonding the bias lines to the PCB. The wafer is held in place with BeCu bipod springs bolted to a copper clamp. Each bias line bond pad on the routing wafer is wire bonded to two pads on the PCB for four-lead measurements with a Lakeshore 372 resistance bridge. The setup for the routing wafer cryogenic screening is in Figure \ref{fig:wafer_screener}. Measurements from the cryogenic testing are used to ensure acceptable shunt resistor values and critical current limits. Additionally, the bias lines are probed for shorts again at 100\,mK, since some shorts detected at 300\,K will disappear at low temperatures (consistent with the apparent shorts being due to conduction through impurities in silicon). 

\subsection{Assembly alignment and clamping}

The next step after screening is to bolt the thin tray onto a copper jig, which is used for subsequent assembly steps and cold testing. The routing wafer is then positioned on the thin tray and aligned with two 1.5\,mm dowel pins. The alignment of the silicon parts within the UMM package is critical for two reasons: (1) limiting offsets between the $\sim$200\,$\mu$m wide bond pads, which would limit automation efficiency and can degrade the strength of the bonds, and (2) ensuring that the routing wafer is centered on the thin tray, lest the thermal contractions during cooling to 100\,mK break wire bonds or the wafer itself.

\begin{figure} [t]
   \begin{center}
   \begin{tabular}{c} 
   \includegraphics[height=8cm]{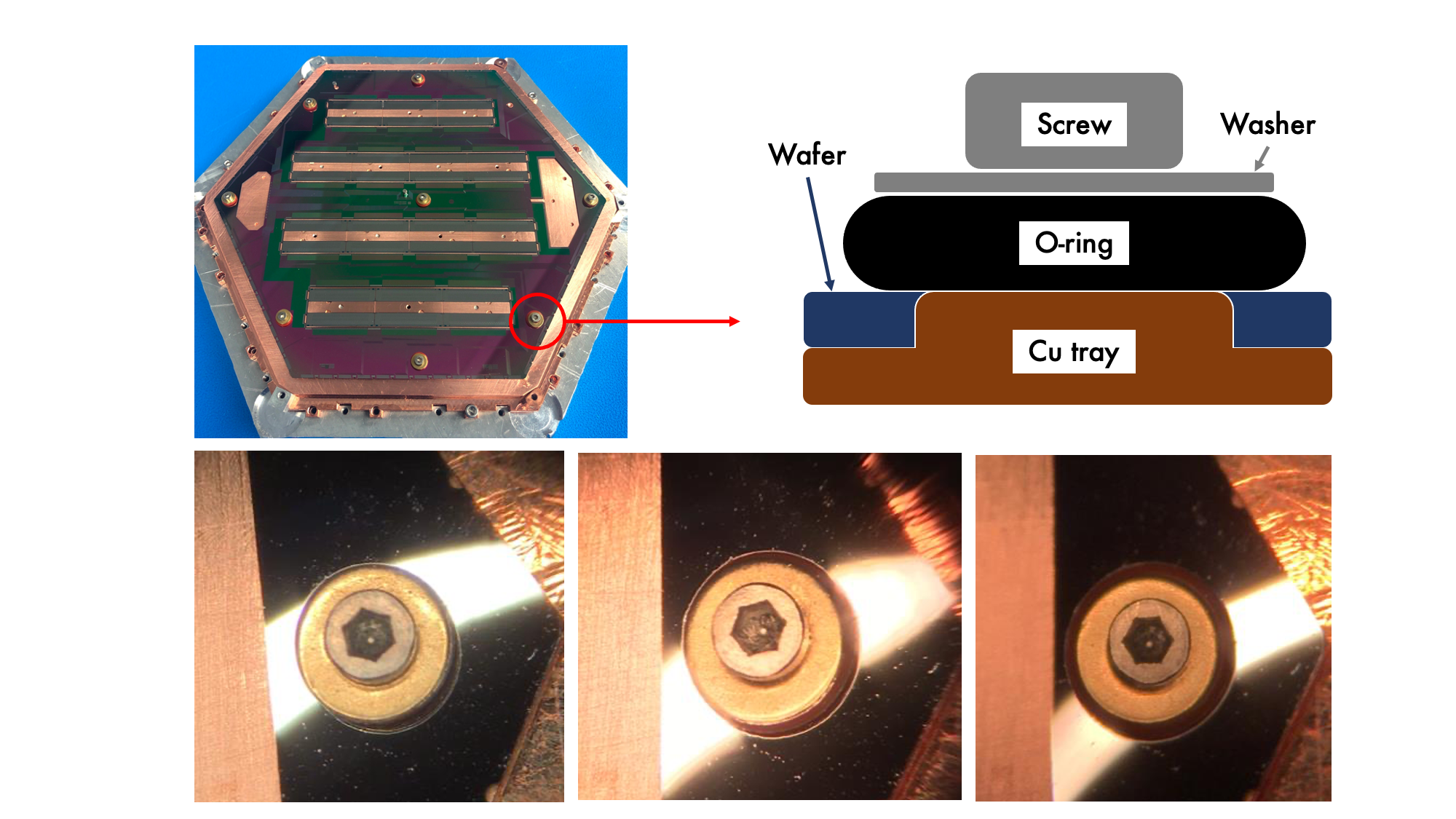}
   \end{tabular}
   \end{center}
   \caption[example] 
   { \label{fig:o-ring} 
   Clamping scheme for UMM assembly. \textbf{Upper left}: A photo of the routing wafer clamping with o-rings. \textbf{Upper right}: A schematic of the setup. The routing wafer is clamped at nine locations to ensure minimal movement during wire bonding. \textbf{Lower left}: Photo of an under-compressed o-ring, in which the wafer is not clamped at all. \textbf{Lower center}: The o-ring is properly compressed, providing sufficient clamping force to keep the wafer in place for wire bonding. \textbf{Lower right}: The o-ring is over-compressed, as it provides no additional clamping force and is overly stressing the wafer.
}
   \end{figure}

When the routing wafer is properly aligned in the thin tray, it is clamped in place using o-rings compressed by a washer and screw, which bolts into the tray. The outer diameter of the compressed o-ring is used as a proxy for clamping strength. This clamping strategy ensures that the wafer is fully constrained during wire bonding, which, like alignment, is required for the automation of wiring bonding and achieving adequate bond strength. A photo and schematic of the o-ring clamping, as well as a comparison of a properly compressed o-ring versus an over- or under-compressed o-ring, can be see in Figure \ref{fig:o-ring}. 

The mux chips are glued with rubber cement using a Tresky flip-chip bonder, which has 2\,$\mu$m precision. The chips are positioned using alignment marks in the routing wafer. The use of rubber cement as an adhesive enables the replacement of mux chips that develop electrical issues, e.g. shorts to ground. These issues are rare but especially problematic since a short on one mux chip shorts out the entire multiplexing chain.

The DC PCB, which has bond pads for connection to the routing wafer along one edge and a connector for a flexible cable on the other side, is bolted to the thin tray. The RF PCB is mated to the thin tray with silver epoxy for grounding purposes and bolted for additional mechanical robustness. These steps are completed before wire bonding for ease of electrical probing during later assembly steps.

\subsection{UMM wire bonding}
\begin{figure} [t]
   \begin{center}
   \begin{tabular}{c} 
   \includegraphics[height=9cm]{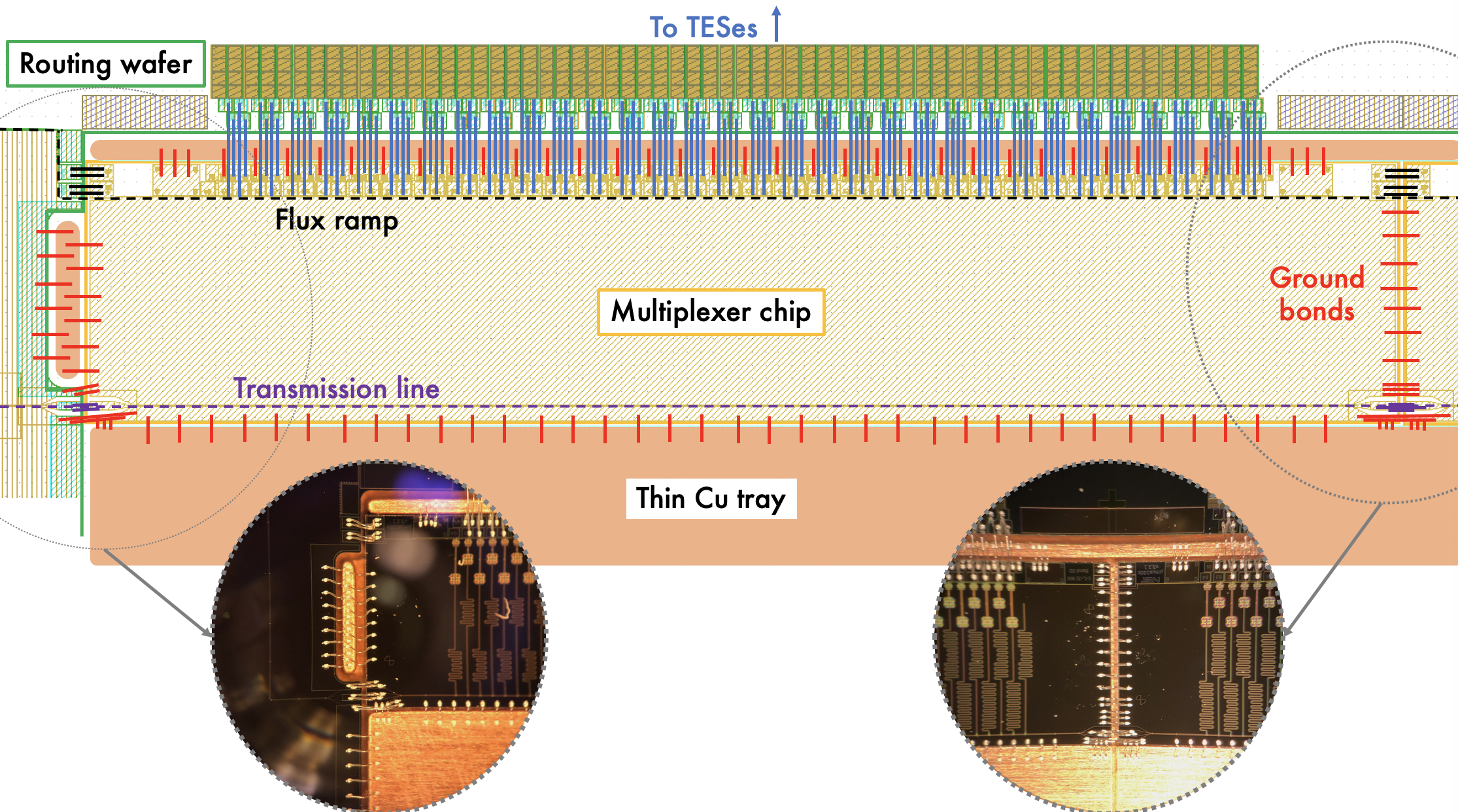}
   \end{tabular}
   \end{center}
   \caption[example] 
   { \label{fig:mux_chip_bonds} 
   The wire bonds for each multiplexer chip. Dense RF ground bonds to the copper tray surround each chip. Wire bonds also couple the multiplexer chip to the transmission line and flux ramp line. Resonator channels are bonded from the chip to the routing wafer, which routes the channels to the individual TESes. \textbf{Left insert}: Photo of wire bonds between chip, wafer, and thin tray.  \textbf{Right insert}: Photo of wire bonds between two chips.
}
   \end{figure}

There are multiple types of RF and DC wire bonds to the multiplexer chips. Mux chips have dense (3 per mm) ground bonds around their perimeter directly to the copper tray and neighboring mux chips. Additional highly densely packed ground bonds flank the transmission bonds at the CPW launches. The lengths of these ground bonds and the CPW bonds are minimized to mitigate their impedance costs while long enough to accommodate relative movement between components during cryocycling. The RF performance of these bonds in SO multiplexing devices has been studied and reported on Li, et al.\cite{Li}. The routing wafer is also grounded directly to the thin tray at the ends of rows of mux chips. Flux ramp and TES channel bonds are placed between mux chip and routing wafer, though their length requirements are less stringent, as they carry low frequency and DC signals, respectively. The wire bonds for a single mux chip can be seen in Figure \ref{fig:mux_chip_bonds}.

In total, each UMM has $\sim$8,000 wire bonds. To scale this assembly step for 49 UMMs, all the wire bonding is programmed for automated bonding on a Delvotec wire bonder. At present, the wire bonding takes roughly 20 hours per assembly, including time for repairing failed or damaged wires, though we expect to improve the speed with efficiencies in the programming.

\subsection{Electrical validation}

During the wire bonding assembly step, we are also routinely checking for shorts and opens. Opens may develop when bonds are missing or fail, both of which can be readily identified with microscope inspection and probing. Shorts, however, can be much more difficult to track down. For this reason, we have developed several short checking tools, including a simultaneous short checking probing box and a program written to triangulate the suspected location of an electrical short. 

\begin{figure} [t]
   \begin{center}
   \begin{tabular}{c} 
   \includegraphics[height=8cm]{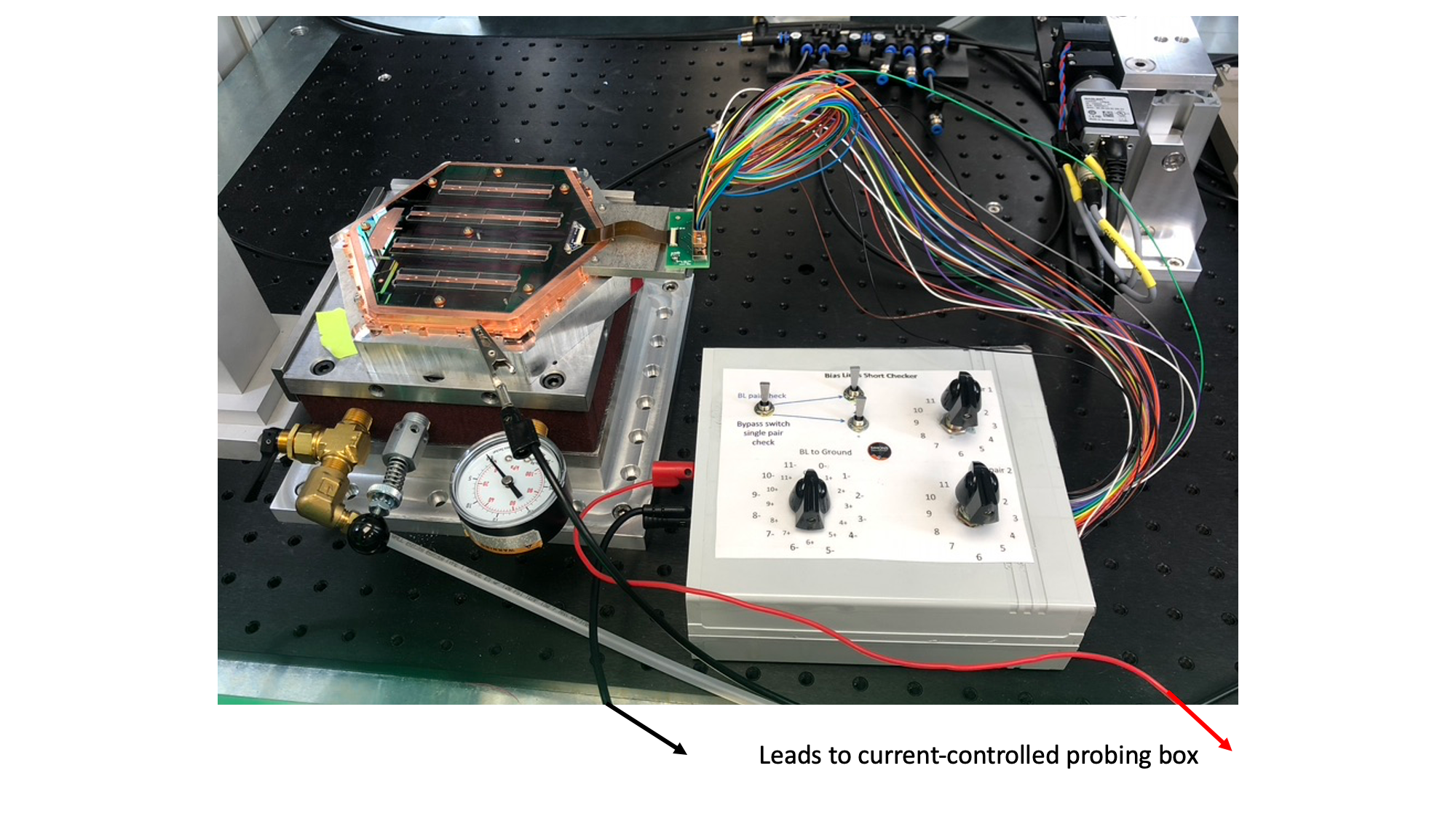}
   \end{tabular}
   \end{center}
   \caption[example] 
   { \label{fig:short_checker} 
   Setup for simultaneously probing for electrical shorts between bias lines while wire bonding. UMM assembly to the left. The box to the right has a series of dials to select which combination of bias lines to probe. Leads from this box plug into a current-controlled probing station.
}
   \end{figure}

Though shorts due to fabrication errors are detected with the initial screening described in Section \ref{intake}, each detector channel has a trace segment that is electrically isolated until the wire bonding step. For this reason, we have developed a short checking protocol employed during wire bonding. This simultaneous short checking box, which is connected to a current-controlled probing station, has dials for selecting which combinations of bias lines are measured. As the automated wire bonding is running, the operator can manually cycle through combinations to quickly identify any insidious shorts that develop between bias lines. The setup for simultaneous short checking is shown in Figure \ref{fig:short_checker}. When shorts between bias lines are detected, the wire bonds to the shorted channel are removed. Assemblies typically have 0---1\% of TES channels left unbonded due to this type of short.

The other important short checking tool, called the universal shorts diagnostic code (USDC), takes advantage of the high resistivity of niobium at 300\,K. Because of the high resistance of the electrical lines, ``shorts" due to wire bonds, or other metal (e.g. wafer traces) in contact, are measured to be $\mathcal{O}$(100) kOhms. Using the resistance measurements between shorted bias lines, the location of a short can be estimated by solving a system of equations. The USDC is also used to triangulate bias line, flux ramp, and CPW shorts to ground.

\subsection{Final packaging}

When wire bonding is complete and the integrity of the electrical lines is validated, the assembly is ready for the final component, the copper lid, to be added. In the final configuration, six spring-loaded pogo pins, glued with Stycast 2850 to the lid, provide the clamping force on the routing wafer. Both the o-rings and dowel pins must be removed to avoid over-constraining the routing wafer during cryocycle. As an intermediary step, BeCu tabs at the wafer corners are added to constrain the wafer while the o-rings and pins are removed. The lid is then bolted to the thin copper tray along surfaces between rows of mux chips. Due to the proximity to the mux chips, brass screws are used to avoid magnetic interference. When the lid is fully integrated, the BeCu corner clamps are removed.

\begin{figure} [t]
   \begin{center}
   \begin{tabular}{c} 
   \includegraphics[height=5.5cm]{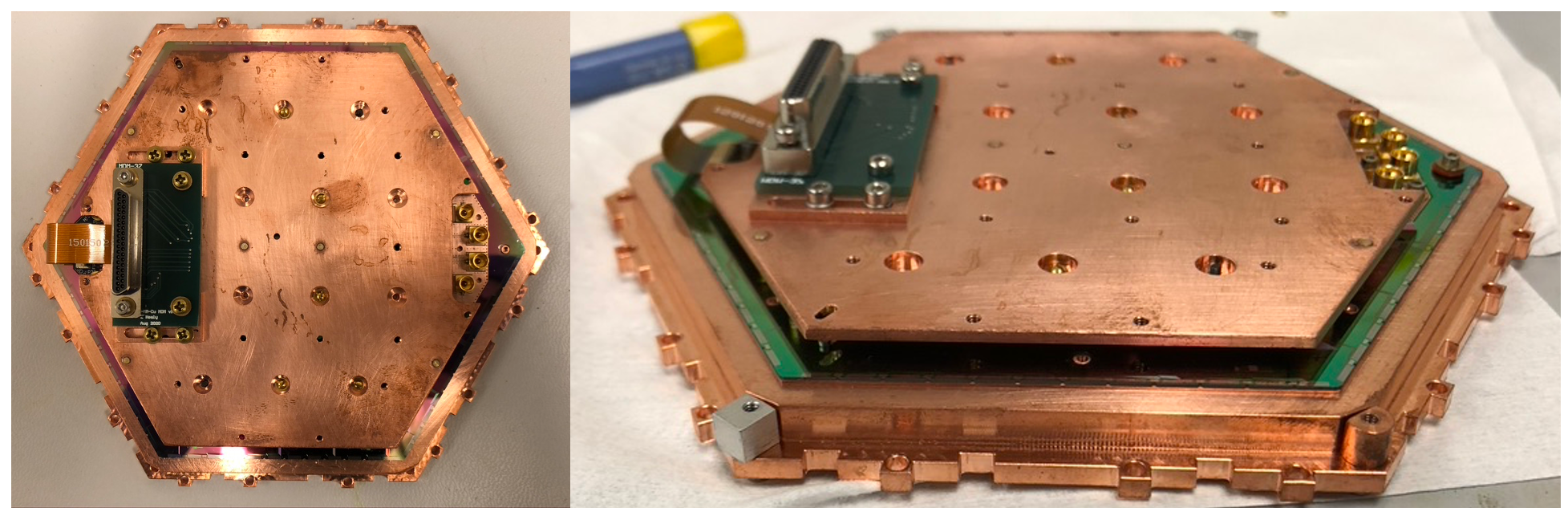}
   \end{tabular}
   \end{center}
   \caption[example] 
   { \label{fig:UMM} 
  Photos of a fully packaged UMM. The package is bolted to a copper jig, used during wire bonding and testing.
}
   \end{figure}

The last step of packaging is to add an interface PCB, which converts the DC connection from a flexible cable to an MDM connector (compatible with optics tube wiring). The thin tray remains bolted to the copper jig for testing. A photo of a fully packaged UMM is in Figure \ref{fig:UMM}.

\section{Testing}

When UMM assembly is complete, the package is tested in a testbed dilution refrigerator to confirm it meets mechanical, thermal, and electrical performance specifications. The testbed readout setup mirrors the SO optics tube design (Figure \ref{fig:umux_ufm_smurf}) with minor modifications\cite{Rao}. Using SMuRF and a vector network analyzer (VNA), data are taken to check for transmission, isolation, noise, and resonator depths and quality factors. Results of the tests have provided a feedback loop for the assembly development.

\section{UFM integration} \label{UFM_integration}

After the performance and robustness of a fully assembled UMM is validated, the readout package is coupled to a detector array. First, the UMM is unbolted from the copper jig. It is then transferred to the top of the detector stack and aligned under microscope. An additional 3,500 wire bonds are placed between the routing wafer and TES detector wafer to connect TESes to their resonator channels. Gold bonds are also laid from PdAu pads on the detector wafer to the gold-plated aluminum feed horns\cite{simon} to create a thermal path off the detector wafer to the 100\,mK bath. The final step of the UFM integration includes adding a mechanical lid for clamping on the stack-up of the detectors and UMM components, necessary for good thermal contact as well as mechanical stability during cool downs and shipping. We have begun testing prototype UFM assemblies and will report on the details of that assembly when those procedures have been fully validated.

\section{Conclusions}

The assembly steps outlined here have been developed to meet SO performance specifications for readout noise and multiplexing factor, as well as cryomechanical and electrical robustness. Additional assembly steps are currently under development to shorten production times and further ensure repeatability between assemblies. One improvement will be validating the thin trays with a high precision coordinate-measuring machine (CMM) before assembly integration. We are also working on an upgrade to the simultaneous short checking setup, which would automate checking for shorts between bias lines during bonding steps. 

To date, we have built two deployment-grade devices, capable of coupling to a detector array. The design of both UFMs and UMMs will also need additional attention in a few areas, which could influence assembly steps. For example, the UMM lid could still be optimized for further reduction of loss due to box mode coupling by testing different materials and geometries. Another important study is the magnetic shielding of the multiplexing modules, which is ongoing \cite{Eve}.The testbed magnetic shielding differs from the field environment, so further investigation will be necessary and may result in the addition of magnetic shielding components. Similarly, the cabling in the testbed refrigerators allows sufficient strain relieving of the wires near the focal plane modules, but we plan to develop improved strain relieving directly to the outer package of the UFM, mitigating any risk of damage during installation. These improvements will ensure that the fielded UMM devices will meet the SO performance goals.

\acknowledgments 
None of this work would be possible without Bert Harrop and Martina Macakova, who have been instrumental in developing and executing the assembly techniques and protocols described here. 

Zhilei Xu is supported by the Gordon and Betty Moore Foundation

\bibliography{SPIEbib}
\bibliographystyle{SPIE} 

\end{document}